# Stacking order in graphite films controlled by van der Waals technology

*Yaping Yang[1,2], Yi-Chao Zou[3], Colin Robert Woods[1,2], Yanmeng Shi[1,2], Jun Yin[1], Servet Ozdemir[1], Takashi Taniguchi[4], Kenji Watanabe[4], Konstantin Sergeevich Novoselov[1,2], Sarah Jane Haigh[3], Artem Mishchenko[1*]*

[1]*School of Physics and Astronomy, University of Manchester, Oxford Road, Manchester, M13 9PL, UK*
[2]*National Graphene Institute, University of Manchester, Oxford Road, Manchester, M13 9PL, UK*
[3]*School of Materials, University of Manchester, Manchester M13 9PL, UK*
[4]*National Institute for Materials Science, 1-1 Namiki, Tsukuba, 305-0044, Japan*
*\*e-mail: artem.mishchenko@gmail.com*

**In graphite crystals, layers of graphene reside in three equivalent, but distinct, stacking positions typically referred to as A, B, and C projections. The order in which the layers are stacked defines the electronic structure of the crystal, providing an exciting degree of freedom which can be exploited for designing graphitic materials with unusual properties including predicted high-temperature superconductivity and ferromagnetism. However, the lack of control of the stacking sequence limits most research to the stable ABA form of graphite. Here we demonstrate a strategy to control the stacking order using van der Waals technology. To this end, we first visualise the distribution of stacking domains in graphite films and then perform directional encapsulation of ABC-rich graphite crystallites with hexagonal boron nitride (hBN). We found that hBN-encapsulation which is introduced parallel to the graphite zigzag edges preserves ABC stacking, while encapsulation along the armchair edges transforms the stacking to ABA. The technique presented here should facilitate new research on the important properties of ABC graphite.**

The possible atomic stacking arrangements for the basal planes in *N*-layer graphite films encompasses $2^{N-2}$ different stacking sequences[1]. The two limiting cases are Bernal stacking (ABA) and the rhombohedral stacking (ABC), with very distinct electronic band structures[2-5]. The ABC-stacked graphite has attracted significant attention because of the promise for fascinating electronic properties. For instance, ABC trilayer, which was the main focus of research on ABC-stacked allotrope, was shown to have a semiconducting behaviour with a tunable band gap[6, 7], insulating quantum Hall states[8, 9], the Lifshitz transition induced by a trigonal warping[10, 11], and the presence of chiral quasiparticles with cubic dispersion[8, 10]. The band structure of bulk rhombohedral graphite hosts 3D Dirac cones, which are gapped out in finite-*N* ABC graphite films[12], thus uncovering topologically protected surface states with nearly flat band dispersions[12, 13]. These nearly flat bands at the surfaces are conducive to strongly correlated phenomena, giving rise to states with spontaneously broken symmetries, such as magnetic ordered states[14, 15] and surface superconducting states[16, 17]. Despite the promise of interesting physics, there are no experimental works on electronic transport in ABC-stacked graphite films thicker than tri- or tetralayer. This is due to the difficulty of producing the ABC allotrope on demand and of high quality. Here, we show that stacking order can be manipulated during the micromechanical assembly of graphite-hBN heterostructures, allowing the production of high-quality ABC-stacked graphite films.

Using bulk natural graphite crystals as a starting material, we exfoliated graphite films onto oxidised silicon substrates and used Raman spectroscopy to identify the presence of ABC stacking[18] (for optical transparency we limited the thickness of flakes to 10 nm). The optical micrograph of one of the exfoliated flakes (Flake1) is shown in Figure 1a. Despite the uniform thickness (≈9 nm or 27±1 graphene layers) and nearly featureless topography, as shown by the atomic force microscopy (AFM) in Figure 1c, the Raman map of 2D peak bandwidth exhibits three regions with strikingly different contrast, Figure 1b. According to previous reports[19-23], these distinct regions should arise from the different stacking order in graphite. To further understand the origin of these three regions, we probed the detail of the Raman spectrum of each region, as shown in Figure 1d. In regions I and III, the 2D band shows the line shape characteristic of ABA and ABC stacking, respectively[19, 23]. Also, the G band of region III is ≈1 cm$^{-1}$ red-shifted compared to that of region I, which is also characteristic of ABC stacking[20, 23]. For region II, the line shape of the 2D band displays a shape characteristic of mixed ABA and ABC stacking. Specifically, the intensity of the low-frequency shoulders of the 2D band (one ranging 2600-2650 cm$^{-1}$ and another ranging 2675-2705 cm$^{-1}$) in region II is weaker than that in region III, while stronger than that in region I. Besides 2D and G modes, we also found a distinct behaviour of regions I, II, and III in the intermediate frequency range (multiple modes between 1650 cm$^{-1}$ and 2500 cm$^{-1}$), see insets in Figure 1d and Supplementary Information (SI) for details.

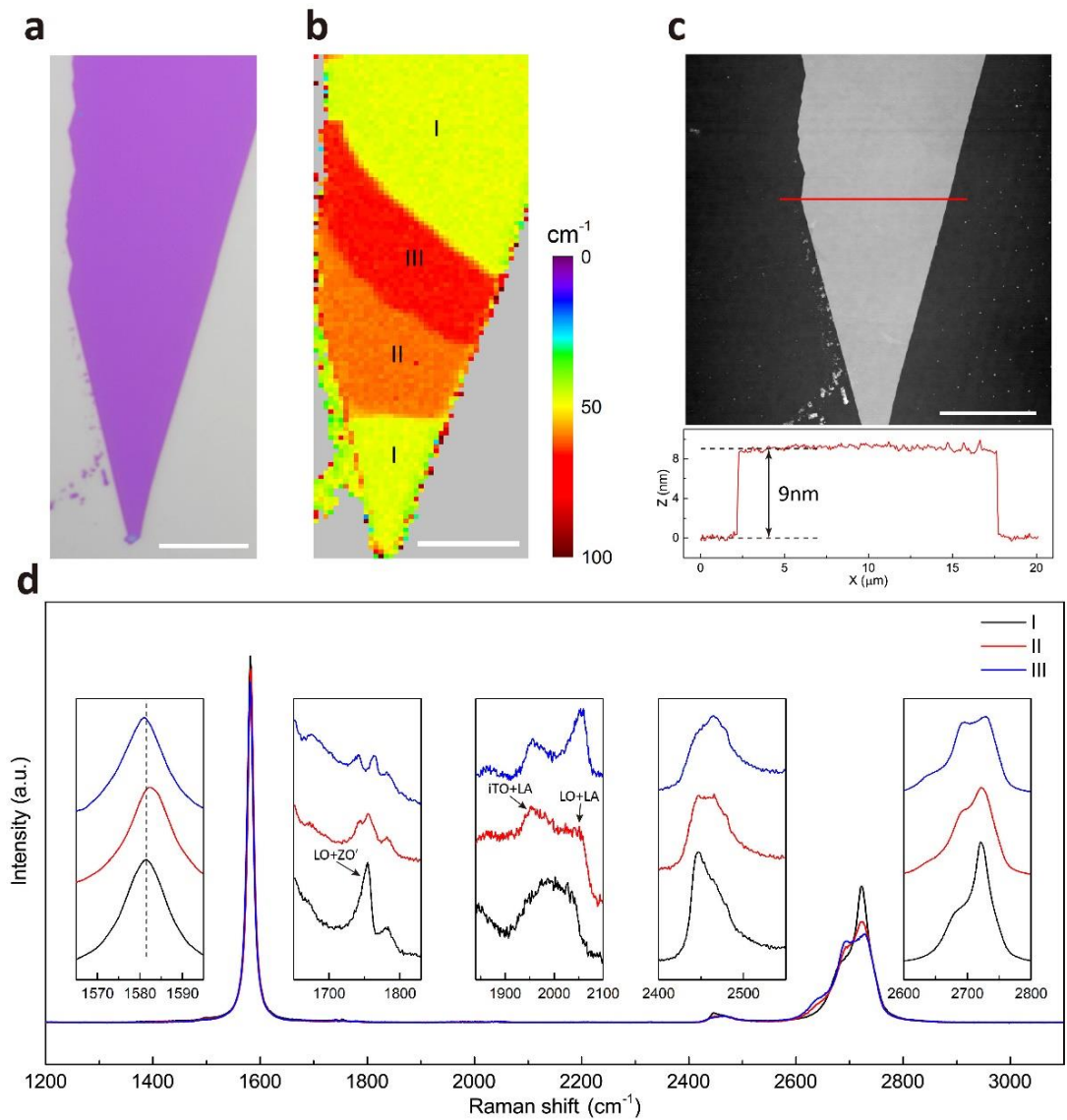

**Figure 1.** Raman spectroscopy of 9-nm-thick graphite flake with different stacking domains. (a,b) Optical micrograph (a) and Raman map (b) of the 2D band width of the flake. The step size of the map is 0.5 μm. (c) AFM image and height profile of the flake. The scale bars in (a-c) are 10 μm. (d) Raman spectra of different regions shown in (b). Insets are the magnified spectra of the G band (1565-1595 cm$^{-1}$), intermediate frequency modes (1650-1830 cm$^{-1}$ and 1840-2100 cm$^{-1}$), G* band (2400-2550 cm$^{-1}$), and 2D band (2600-2800 cm$^{-1}$). G band corresponds to the high-frequency E$_{2g}$ phonon at the Γ point. The intermediate frequency modes are related to out-of-plane layer breathing phonons. Both G* band and 2D bands originate from the intervalley double resonance process. All these modes are sensitive to layer number and stacking order.

In our further study we have focused on 2D and G bands, since they have strong signals and both contain information about the stacking order. We observed a remarkable richness of different domains with distinct Raman scattering responses in another exfoliated graphite film (Flake2), presented in Figure 2. We focused on the region encompassed by the dashed line in Figure 2a,c,d which shows homogeneous thickness (see also Figure S1 in Supporting Information). According to Refs[19, 23] and supported by our measurements of the 2D band of Flake1, the relative intensity of the lower frequency component (≈2680 $cm^{-1}$) to the higher frequency component (≈2725 $cm^{-1}$) changes with the stacking order. Thus, for Flake2 we mapped the ratio of the integral area of the 2D band ranging 2675-2705 $cm^{-1}$ (indicated by blue rectangle in Figure 2b) to the area in the range 2705-2735 $cm^{-1}$ (indicated by grey rectangle in Figure 2b). In the ratio map, plotted in Figure 2a, we can identify 7 regions with distinct ratios, indicating at least 7 domains with different stacking orders. The Raman maps of 2D band width and position (Figures S2a and b in Supporting Information) also show consistent shape and distribution of the domains. The 2D mode (a D mode overtone) arises from a double resonance process, which makes it sensitive to electronic band structure, and thus, to the stacking sequence[24, 25]. This suggests that the evolution of the 2D band line shape in regions 1 to 7 in Figure 2 reflects the increasing proportion of ABC stacking; from perfect ABA (region 1) to perfect ABC stacking (region 7).

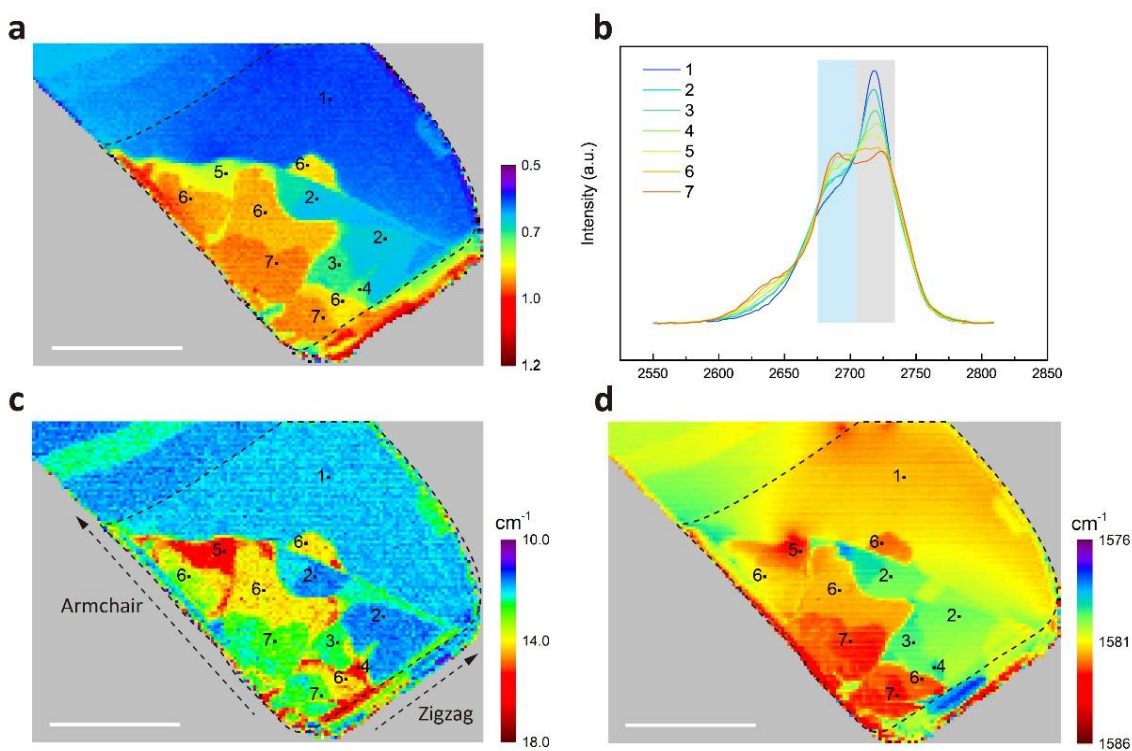

**Figure 2.** The evolution of 2D and G bands among different domains in 6.5-nm-thick graphite flake. (a) Raman map of the ratio of the integral area of 2D band ranging 2675-2705 $cm^{-1}$ (indicated by blue rectangle) to that ranging 2705-2735 $cm^{-1}$ (indicated by pink rectangle). (b) 2D band regions corresponding to domains 1 to 7 shown in (a). (c,d) Raman maps of the G band width and position. The two orthogonal black dashed arrows in (c) indicate the edge chirality of the flake, confirmed by the D band intensity map (Figure S2c in Supporting Information). The scale bars are 20 μm.

The G mode originates from doubly degenerate optical phonons at the centre of the Brillouin zone, and is not affected by the electronic structure – the frequency and width of the G band reflect the changes in phonon dispersion due to the difference in the stacking sequence[19, 26]. In the G band width and position maps (Figures 2c and d), the distribution and shape of the domains match those in the ratio map and 2D band width map (Figure 2a and Figure 2Sa in the Supporting Information), also confirming the different stacking order of these domains. However, the trend of G band width and position among these domains are slightly different from the ratio map (Figures 2a), which we attribute to the influence of local strain[27]. We noticed three special regions where local mechanical deformation of the flake allows us to investigate the effect of the strain. There is a wrinkle diagonally across Flake2 (seen in dark field image and AFM in Figure S1 in the Supporting Information), where an out of plane crease leads to increased strain in the graphite. For region 1, the G band position gradually decreases along the direction from the top of the flake to the edge of the wrinkle. Similar behaviour can also be seen in 2D band position map (Figure S2b in the Supporting Information). For region 5, where the wrinkle vanishes, and therefore the strain should also be reasonably high, both the G band width and position are higher than that of the adjacent regions. For region 7, there is a contamination bubble between the two domains (shown in Figure S1a in the Supporting Information), and thus the strain induced broadening and softening of the G band also occur near the bubble.

Diffraction in the transmission electron microscope (TEM) allows to measure the fraction of ABA and ABC stacking from the relative intensity of the first order diffraction peaks, and cross-sectional TEM imaging is able to directly show the local stacking order in different domains (shown in Figure 3 and Figure S8 in the Supporting Information). To further examine the stacking order present in plan-view ABC/ABA graphite specimens, selected-area electron diffraction (SAED) and dark field imaging (Figure 3) were used. For ABC graphite the intensity of the first-order diffraction peaks is close to zero, whereas for ABA stacking the first-order diffraction peaks are more prominent.[28] Figure 3a shows an optical image taken from a typical ABC/ABA graphite flake (Flake3, ≈6 nm thick) which was transferred onto the TEM grid, the corresponding 2D Raman map is shown in Figure 3b. Regions 1[#] and 2[#], ABC and ABA stacking respectively, were further examined by electron diffraction, and the result is shown in Figures 3d and e. The fraction of ABC stacking can be evaluated by calculating the intensity ratio (R) between the six inner-most first-order $\{10\bar{1}0\}$ spots and the six second-order $\{11\bar{2}0\}$ spots[28] (see the Supporting Information). The calculated fraction of ABC for regions 1[#] and 2[#] are 96% and 2%, respectively, in excellent agreement with the Raman analysis.

Figure 3c shows the TEM dark field images performed with the $(10\bar{1}0)$ selected diffraction peak, where one can easily distinguish between the ABA and ABC stacked domains. The greater intensity in the dark field for region 2[#] indicates the presence of a high level of ABA stacking consistent with the Raman analysis. The holey $SiN_x$ support also appears bright in the dark field image due to its large thickness. It is interesting to see that some ABC/ABA mixing regions (e.g. region 3[#]) can still be preserved after the flake transfer according to the Raman map (Figure 3b). Figure 3f shows a diffraction pattern taken from region 3[#] and the calculated $F_{ABC}$ is 81%, confirming a mixed region of ABC and ABA stacking. Such TEM analysis is too time consuming to measure large numbers of flakes but the correlation between our Raman and TEM analysis provides excellent calibration for large Raman data sets.

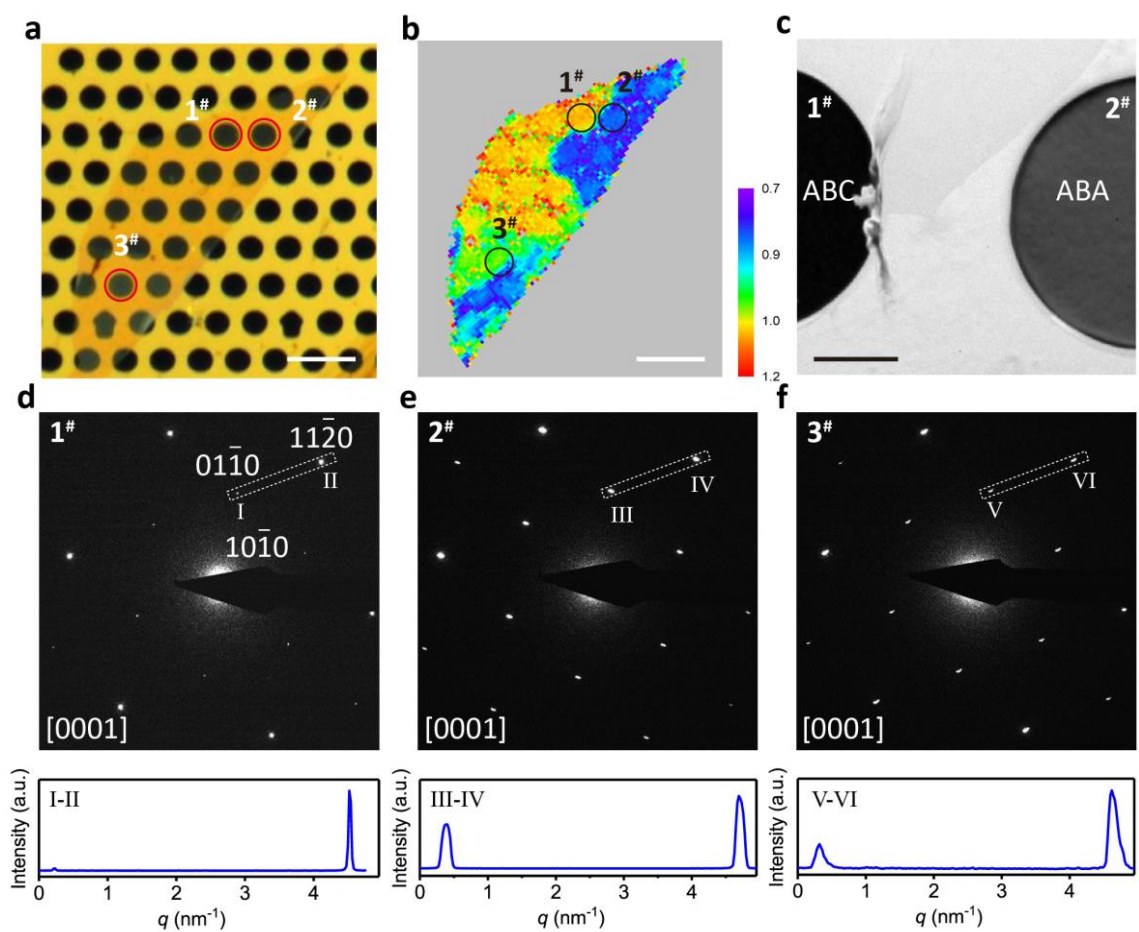

**Figure 3.** Plan-view TEM imaging and diffraction of an ABC/ABA graphite flake. (a) Optical micrograph of Flake3 suspended on a TEM grid, (b) Corresponding Raman map of the ratio of the integral area of 2D band ranging 2675-2705 cm$^{-1}$ to that ranging 2705-2735 cm$^{-1}$. (c) Dark field TEM image showing the different domains in holes 1$^\#$ and 2$^\#$ obtained using the intensity from the (10$\bar{1}$0) diffraction peak which is much stronger for ABA domains. (d,e) SAED taken from region 1$^\#$ and 2$^\#$ and beneath these the line intensity profile measured along line I-II and III-IV respectively, allowing comparison of the intensity of the first order 10$\bar{1}$0 and secod order 11$\bar{2}$0 diffraction spots. (f) SAED patterns taken from region 3$^\#$ with the bottom inset showing the line intensity profile taken along line V-VI. Scale bars, 10 μm for panels (a) and (b), 1 μm for panel (c).

We exfoliated 617 flakes in total, 176 of them contained an ABC-stacked region (29%). We did not observe any effect of the tape peel direction on the success rate of production of ABC-stacked flakes. Among the flakes with ABC-stacked regions that have been mapped (24 in total), 21 had different domains (88%). We found that domains with different stacking sequences in graphite films also have a distinct mechanical response and exhibit marked contrast in near-field infrared nanoscopy imaging, as shown in Figures S4 and S5 in the Supporting Information. Generally, rhombohedral stacking in graphite is quite rare as it is energetically less stable compared to Bernal stacking – it was reported that in the exfoliated samples around 15% of the total area displays rhombohedral stacking[19]. However, our results demonstrate that among almost all the graphite films containing rhombohedral stacking regions, there exist domains with various proportions of rhombohedral stacking segments. The formation of these domains is most likely related to the slip-induced transformation of stacking order during micromechanical exfoliation, due to the unavoidable presence of shear force applied to graphite crystallites through a soft polymer stamp used for exfoliation (see Supporting Information for details).

A shear force applied to graphite leads to a displacement of the graphene planes in a 'stick-slip' fashion[29]. If the displacement occurs along zigzag edge directions, it preserves the stacking order, while displacement parallel to armchair edges alternates stacking order between ABA and ABC. This effect is illustrated in Figure 4a (using ABC trilayer as an example); all the stacking orders of trilayer graphene after the relative movement of the top graphene layer with different displacement vectors are summarised in Table S1 in the Supporting Information. The shear force along the armchair directions causes the movement of the graphene layers, leading to a transition of the stacking order. These shear-induced basal plane dislocations cause the graphene layers to wrinkle and kinks to arise[30]. This coincides with the fact that samples with wrinkles always appear with different domains in agreement with the reported result that the formation of kink bands can induce local stacking faults in layered crystals[31]. As the number of graphene layers increases, it is harder for the thicker flakes to fold since they have larger bending moduli, thus the need to induce stacking faults to accommodate bending, which explains the formation of the observed domains in graphite films[31]. The domain size is governed by a balance between stacking fault energy and dislocation energies, and the dislocations could be splitting into lower energy partial dislocations which have stacking faults between them[32]. Interestingly, the number of observed different domains remains rather low. For instance, the sample in Figure 2 is approximately 18-layer-thick, which could lead to tens of thousands of domains with distinct stacking orders covering all the spectrum between pure ABA and pure ABC types. But just around seven were observed.

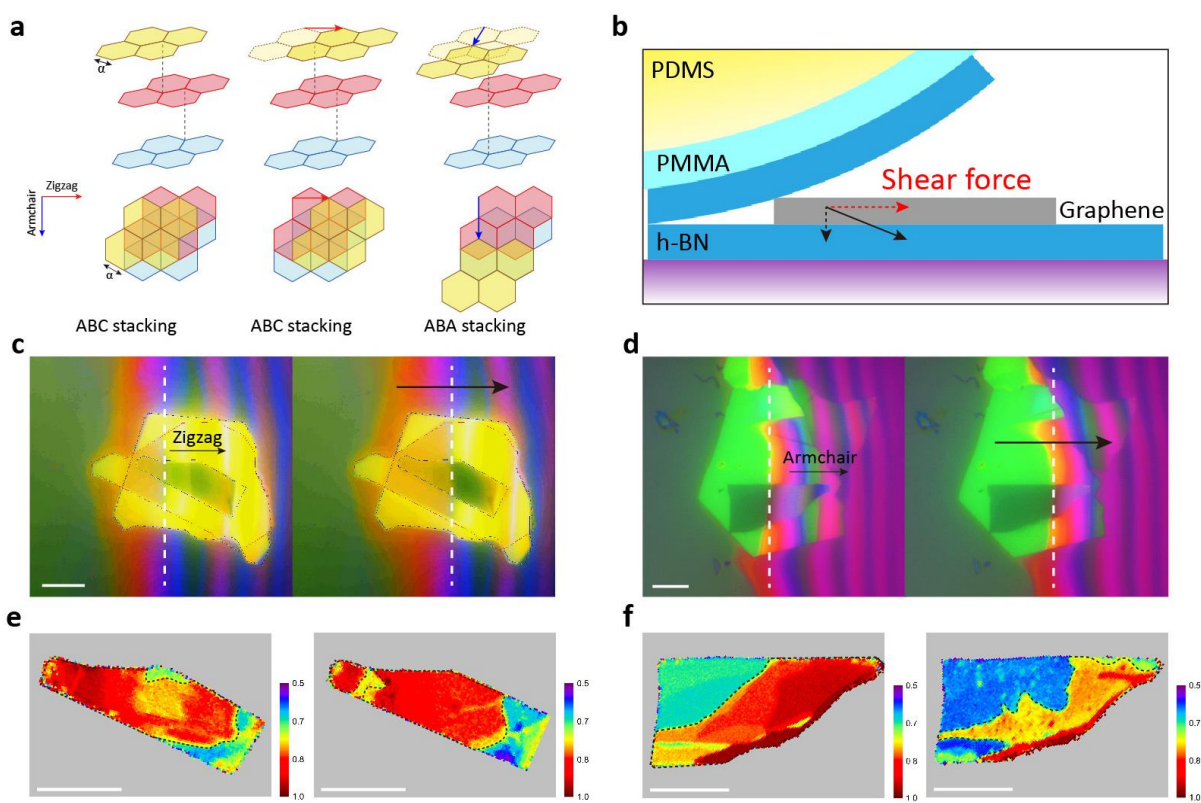

**Figure 4.** Stacking order control in graphite films. (a) Crystal structure and top view of the stacking order transformation of a model trilayer graphene. The top, middle, and bottom layers are labelled yellow, red, and blue, respectively. The left figure is the original ABC stacking trilayer. The middle and right figures are the final stacking order after the energy favourable displacement of the top layer along zigzag (red arrow) and armchair (blue arrow) directions, respectively. α is the C-C bond length. (b) Schematic of the shear force applied to the flakes during micromechanical transfer. (c,d) Optical images of the pressing down process of PDMS during transfer along zigzag (Flake 4) or armchair (Flake 5) directions, respectively. The edge of the contact area is marked by white dashed line. The black arrows indicate the expansion direction of the edge. (e,f) Maps of the ratio of the integral area of 2D band ranging 2675-2705 cm$^{-1}$ to that ranging 2705-2735 cm$^{-1}$ before (left) and after (right) encapsulation for Flake 4 and 5, respectively. The outline of the ABC stacking region is marked by the black dashed line. The scale bars are 20 μm.

Inspired by this mechanism of stacking transformation, we modified the encapsulation protocol for ABC-stacked graphite, aiming to significantly increase the chance of the survival of this energetically less stable stacking allotrope. We used the PDMS (polydimethylsiloxane) transfer technique[33], in which viscoelastic PDMS enables precise control of the contact area between PMMA (polymethylmethacrylate) adlayer and the substrate, including the contact area expansion direction while pressing down PDMS, and the expansion speed. During the transfer process, there is a stress force while pressing down the top h-BN/PMMA/PDMS stack, which can be resolved into two components, the vertical one and the horizontal one that is parallel to the surface of the substrate, denoted as the shear force, as shown in Figure 4b. As discussed above, the transformation of the stacking order is mainly caused by the shear force applied to the exfoliated flakes. If the direction of this shear force is aligned with the zigzag edge direction of graphite crystal, then its stacking order will remain unchanged. Therefore, the key is to distinguish the edge chirality of graphite films. If the angle between two adjacent edges of a graphite film is 30°, 90°, or 150°, the two edges have different chiralities, i.e., one armchair and the other zigzag. This edge chirality could be determined by Raman spectroscopy using the D band intensity. A perfect zigzag edge cannot produce a D peak, whereas the armchair edges contribute to the D band[18, 34]. Thus, the intensity of the D band near the armchair edge is always stronger than that near the zigzag edge[34].

Using this modified van der Waals assembly process, we encapsulated an ABC-stacked graphite film (Flake4) in which the angle between two of its edges is 150° (Figure S6a in the Supporting Information). The D band intensity map tells the edge chirality (Figure S6b in the Supporting Information). During the transfer, we carefully pressed down the PDMS along the zigzag edge in order to align the shear force with the zigzag direction, as shown in Figure 4c. The Raman maps of Flake4 before and after encapsulation (Figure 4e) clearly show that the ABC stacking order was successfully preserved. As a comparison, we also made another encapsulated ABC stacked graphite film (Flake5) with the PDMS pressed down along the armchair edge of the graphite film during the transfer process. The result shows that after transfer the ABC stacking region has almost disappeared (Figure 4f). Our technique proved to be reproducible: we made five encapsulated samples using zigzag shear transfer, and for these flakes, the ABC stacking regions of the graphite films all remained with little change in their shape. Our approach to micromechanical control of the crystal structure of graphite enables preparation of high-quality hBN-encapsulated rhombohedral graphite films and should greatly promote electronic transport studies in this exciting material.

## Acknowledgements
A.M. acknowledges the support of EPSRC Early Career Fellowship EP/N007131/1. S.J.H. and Y.-C. Zou acknowledge financial support from the ERC H2020 Starter Grant EvoluTEM and EPSRC (EP/P009050/1) and assistance in diffraction analysis from Mr. Eupin Tien.

## References

1. Min, H. K.; MacDonald, A. H. *Phys. Rev. B* **2008,** 77, (15), 155416.
2. Mak, K. F.; Shan, J.; Heinz, T. F. *Phys Rev Lett* **2010,** 104, (17), 176404.
3. Bao, C.; Yao, W.; Wang, E.; Chen, C.; Avila, J.; Asensio, M. C.; Zhou, S. *Nano Lett* **2017,** 17, (3), 1564-1568.
4. Koshino, M. *Phys. Rev. B* **2010,** 81, (12), 125304.

5.	Avetisyan, A. A.; Partoens, B.; Peeters, F. M. *Phys. Rev. B* **2010,** 81, (11), 115432.
6.	Lui, C. H.; Li, Z. Q.; Mak, K. F.; Cappelluti, E.; Heinz, T. F. *Nature Phys.* **2011,** 7, (12), 944-947.
7.	Zou, K.; Zhang, F.; Clapp, C.; MacDonald, A. H.; Zhu, J. *Nano Lett* **2013,** 13, (2), 369-73.
8.	Bao, W.; Jing, L.; Velasco, J.; Lee, Y.; Liu, G.; Tran, D.; Standley, B.; Aykol, M.; Cronin, S. B.; Smirnov, D.; Koshino, M.; McCann, E.; Bockrath, M.; Lau, C. N. *Nature Phys.* **2011,** 7, (12), 948-952.
9.	Chen, G.; Jiang, L.; Wu, S.; Lv, B.; Li, H.; Watanabe, K.; Taniguchi, T.; Shi, Z.; Zhang, Y.; Wang, F. *arXiv:1803.01985* **2019**.
10.	Zhang, L.; Zhang, Y.; Camacho, J.; Khodas, M.; Zaliznyak, I. *Nature Phys.* **2011,** 7, (12), 953-957.
11.	Koshino, M.; McCann, E. *Phys. Rev. B* **2009,** 80, (16).
12.	Ho, C. H.; Chang, C. P.; Lin, M. F. *Phys. Rev. B* **2016,** 93, (7), 075437.
13.	Xiao, R. J.; Tasnadi, F.; Koepernik, K.; Venderbos, J. W. F.; Richter, M.; Taut, M. *Phys. Rev. B* **2011,** 84, (16), 165404.
14.	Olsen, R.; van Gelderen, R.; Smith, C. M. *Phys. Rev. B* **2013,** 87, (11).
15.	Pamuk, B.; Baima, J.; Mauri, F.; Calandra, M. *Physical Review B* **2017,** 95, (7).
16.	Kopnin, N. B.; Ijäs, M.; Harju, A.; Heikkilä, T. T. *Phys. Rev. B* **2013,** 87, (14).
17.	Muñoz, W. A.; Covaci, L.; Peeters, F. M. *Phys. Rev. B* **2013,** 87, (13).
18.	Ferrari, A. C.; Basko, D. M. *Nat Nanotechnol* **2013,** 8, (4), 235-46.
19.	Lui, C. H.; Li, Z.; Chen, Z.; Klimov, P. V.; Brus, L. E.; Heinz, T. F. *Nano Lett* **2011,** 11, (1), 164-9.
20.	Cong, C.; Yu, T.; Sato, K.; Shang, J.; Saito, R.; Dresselhaus, G. F.; Dresselhaus, M. S. *ACS Nano* **2011,** 5, (11), 8760-8.
21.	Casiraghi, C.; Hartschuh, A.; Lidorikis, E.; Qian, H.; Harutyunyan, H.; Gokus, T.; Novoselov, K. S.; Ferrari, A. C. *Nano Lett.* **2007,** 7, (9), 2711-2717.
22.	Katsnelson, M. I.; Novoselov, K. S. *Solid State Commun.* **2007,** 143, (1-2), 3-13.
23.	Henni, Y.; Ojeda Collado, H. P.; Nogajewski, K.; Molas, M. R.; Usaj, G.; Balseiro, C. A.; Orlita, M.; Potemski, M.; Faugeras, C. *Nano Lett* **2016,** 16, (6), 3710-6.
24.	Maultzsch, J.; Reich, S.; Thomsen, C. *Phys. Rev. B* **2004,** 70, (15).
25.	Cançado, L. G.; Reina, A.; Kong, J.; Dresselhaus, M. S. *Phys. Rev. B* **2008,** 77, (24).
26.	Yan, J. A.; Ruan, W. Y.; Chou, M. Y. *Phys. Rev. B* **2008,** 77, (12).
27.	Huang, M. Y.; Yan, H. G.; Chen, C. Y.; Song, D. H.; Heinz, T. F.; Hone, J. *Proc. Natl. Acad. Sci. U. S. A.* **2009,** 106, (18), 7304-7308.
28.	Latychevskaia, T.; Son, S.-K.; Yang, Y.; Chancellor, D.; Brown, M.; Ozdemir, S.; Madan, I.; Berruto, G.; Carbone, F.; Mishchenko, A.; Novoselov, K. S. *Frontiers of Physics* **2018,** 14, (1).
29.	Dienwiebel, M.; Verhoeven, G. S.; Pradeep, N.; Frenken, J. W.; Heimberg, J. A.; Zandbergen, H. W. *Phys Rev Lett* **2004,** 92, (12), 126101.
30.	Soule, D. E.; Nezbeda, C. W. *J. Appl. Phys.* **1968,** 39, (11), 5122-5139.
31.	Rooney, A. P.; Li, Z.; Zhao, W.; Gholinia, A.; Kozikov, A.; Auton, G.; Ding, F.; Gorbachev, R. V.; Young, R. J.; Haigh, S. J. *Nat Commun* **2018,** 9, (1), 3597.
32.	Gong, L.; Young, R. J.; Kinloch, I. A.; Haigh, S. J.; Warner, J. H.; Hinks, J. A.; Xu, Z.; Li, L.; Ding, F.; Riaz, I.; Jalil, R.; Novoselov, K. S. *ACS Nano* **2013,** 7, (8), 7287-94.
33.	Uwanno, T.; Hattori, Y.; Taniguchi, T.; Watanabe, K.; Nagashio, K. *2D Materials* **2015,** 2, (4), 041002.
34.	Cancado, L. G.; Pimenta, M. A.; Neves, B. R.; Dantas, M. S.; Jorio, A. *Phys Rev Lett* **2004,** 93, (24), 247401.

# Supporting Information

# Stacking order in graphite films controlled by van der Waals technology

*Yaping Yang[1,2], Yi-Chao Zou[3], Colin Robert Woods[1,2], Yanmeng Shi[1,2], Jun Yin[1], Servet Ozdemir[1], Takashi Taniguchi[4], Kenji Watanabe[4], Konstantin Sergeevich Novoselov[1,2], Sarah Jane Haigh[3], Artem Mishchenko[1*]*

[1]School of Physics and Astronomy, University of Manchester, Oxford Road, Manchester, M13 9PL, UK
[2]National Graphene Institute, University of Manchester, Oxford Road, Manchester, M13 9PL, UK
[3]School of Materials, University of Manchester, Manchester M13 9PL, UK
[4]National Institute for Materials Science, 1-1 Namiki, Tsukuba, 305-0044, Japan
*e-mail: artem.mishchenko@gmail.com

# 1. Experimental methods

## 1.1. Exfoliation and van der Waals assembly

Graphite films were obtained by mechanical exfoliation of bulk graphite (NGS naturgraphit, Graphenium Flakes 25-30mm) onto the SiO$_2$ (290 nm)/Si substrate cleaned by oxygen plasma. During the exfoliation, the shear force mainly originates from peeling off the water-soluble tape. Whether the extra shear force is applied on top of the tape or not did not make a difference in getting ABC stacking flakes.

The transfer process (van der Waals assembly) can be briefly divided into three main steps: (i) The PMMA/PDMS/glass slide was prepared to pick up the h-BN which was mechanically exfoliated onto a Si/SiO$_2$ (290 nm) wafer; (ii) Then we used the h-BN/PMMA/PDMS/glass slide to pick up graphite flakes; (iii) Finally the graphite flake/h-BN was released onto the bottom h-BN on Si/SiO$_2$ (290 nm) wafer by peeling off the PMMA from the top h-BN.

## 1.2. Raman characterisation

The Raman spectra were acquired by Renishaw Raman system with 1800 lines/mm grating, using linearly polarised laser radiation at the wavelength (photon energy) of 532 nm (2.33 eV). The laser power was kept below 3 mW on the sample surface to avoid laser-induced heating. The Raman spatial maps were taken with a step size of 0.5 µm. To extract the peak width, position and intensity of different bands, we fit the spectrum at each pixel in the spatial mapping to a single Lorentzian function. For the intermediate frequency range, the acquisition time was increased to 10 min because of the weak signal.

## 1.3. Atomic Force Microscopy (AFM)

The AFM images were taken by Bruker AFM system. The nanomechanical images were obtained by PeakForce Quantitative Nanoscale Mechanical (QNM) mode. The Scanasyst-Fluid+ cantilever with the silicon tip on nitride lever is used. The spring constant of the cantilever is 0.7 N/m.

## 1.4. Near-field infrared nanoscopy

The near-field infrared nanoscopy imaging was carried out by scattering-type Scanning Near-Field Optical Microscopy (sSNOM), based on tapping mode AFM, using Bruker Anasys nanoIR3 system. The Quantum Cascade Laser with the wavelength of 9090 nm (narrow-band) was focused onto the apex of the tip with Platinum Iridium coating and the resonant frequency of ≈350 kHz. Near-field optical images with spatial resolution better than 100 nm were taken simultaneously with the topography images. The enhanced optical near field interaction between the tip and sample depends on the electronic properties of the sample. Thus ABA- and ABC-stacked domains give different infrared responses due to their different electronic band structures.

## 1.5. Transmission electron microscope diffraction and cross-sectional imaging

For high resolution scanning transmission electron microscope (STEM) imaging, a probe side aberration corrected FEI Titan G2 80-200 S/TEM microscope was used with an acceleration voltage of 200 kV, probe convergence angle of 21 mrad, an annular dark field (ADF) inner angle of 48 mrad and a probe current of ≈80 pA. To ensure the electron probe was parallel to the basal planes, the cross-sectional sample was aligned to the relevant Kikuchi bands of the graphite crystal. The electron diffraction experiment was conducted using the same microscope but operated in TEM mode with an accelerating voltage of 80 kV.

Cross-sectional TEM specimens were prepared from graphite flakes on SiO$_2$/Si substrate covered by boron nitride, using Ga$^+$ focused ion beam (FIB) scanning electron microscope (SEM) (FEI Helios 660). The region of interest (ROI) was deposited with a narrow Pt strap (1 µm thick) to avoid charging and being damaged by the ion beam when creating the cross-sectional lamella. The ROI contains ABA and ABC stacked domains, confirmed by Raman measurement before FIB milling. Cross section lamellae were prepared perpendicular to the armchair or zigzag edges of the graphite flakes, to facilitate orienting the crystal to perform atomic resolution STEM imaging. Then the lamella was cut from the substrate and transferred to a pillar on a specialist OmniProbe™ TEM grid, followed by 30 kV, 16 kV, 5 kV and 2 kV ion beam milling and polishing for electron transparency.

## 2. The distinct behaviour of the three regions of Flake1 in the combination Raman modes (ranging 1650 cm$^{-1}$ to 2500 cm$^{-1}$)

The intensity of the combination modes in the range from 1650 cm$^{-1}$ to 2100 cm$^{-1}$ is very weak, roughly two orders of magnitude smaller than that of the G mode. These modes are related to out-of-plane layer breathing phonons, and have a direct sensitivity to layer number and stacking order.

In the range of 1650-1850 cm$^{-1}$, it is reported that the M band at ≈1750 cm$^{-1}$ is attributed to LO+ZO' (in-plane longitudinal optical phonon and the out-of-plane interlayer breathing phonon) combination mode, based on an intravalley double-resonant Raman scattering process[1-5]. For ABA stacked multilayer graphene thicker than 6 layers, M band appears as a strong peak near 1750 cm$^{-1}$, while for ABC stacked graphene, M band splits into subpeaks with narrower widths as the number of layer increases[1, 6]. Our results in Figure 1d in the main text show consistent fingerprints of M band. In region I and III, M band exhibits the characters of ABA and ABC stacking, respectively, while in region II, M band splits into two subpeaks, displaying the transition line shape between ABA and ABC stacking. In the range of 1850-2100 cm$^{-1}$, there are two combinational modes, the iTO+LA (in-plane transverse optical phonon and longitudinal acoustic phonon) at ≈1950 cm$^{-1}$ and LO+LA at ≈2050 cm$^{-1}$ [5, 7]. From the bottom region I to region III, this band gradually splits into two subbands, corresponding to the two combinational modes, indicating the transition of the ABA stacking to ABC stacking.

In terms of the Raman band at ≈2450 cm$^{-1}$, named G* (or D+D'') band, it is assigned as the combination of one iTO phonon around the Brillouin zone corner K and one LA phonon in the intervalley double resonance process[8, 9]. In our results, G* band shows unique line shape in each region.

Among these Raman bands arising from the double resonance scattering process, there are two main reasons for their sensitivity to the stacking order. On one hand, the electronic structures of different stacking orders result in different values of the phonon wave vectors required to satisfy the double resonance process. On the other hand, the stacking order directly influences the phonon dispersions, thus affects the electron-phonon coupling processes. These stacking-order-dependent modes showing unique line shapes for different regions again verify the different stacking orders among these three regions.

## 3. Optical and AFM images and Raman maps of Flake2

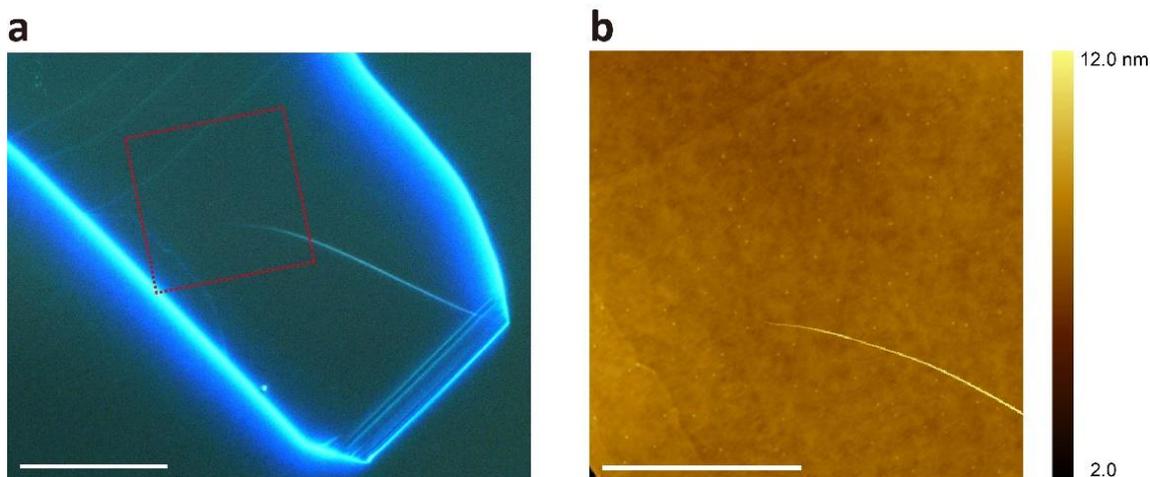

**Figure S1.** Optical and AFM images of Flake2. (a) Optical micrograph, scale bar is 20 μm. (b) AFM image of the dashed area in (a). The scale bar is 10 μm.

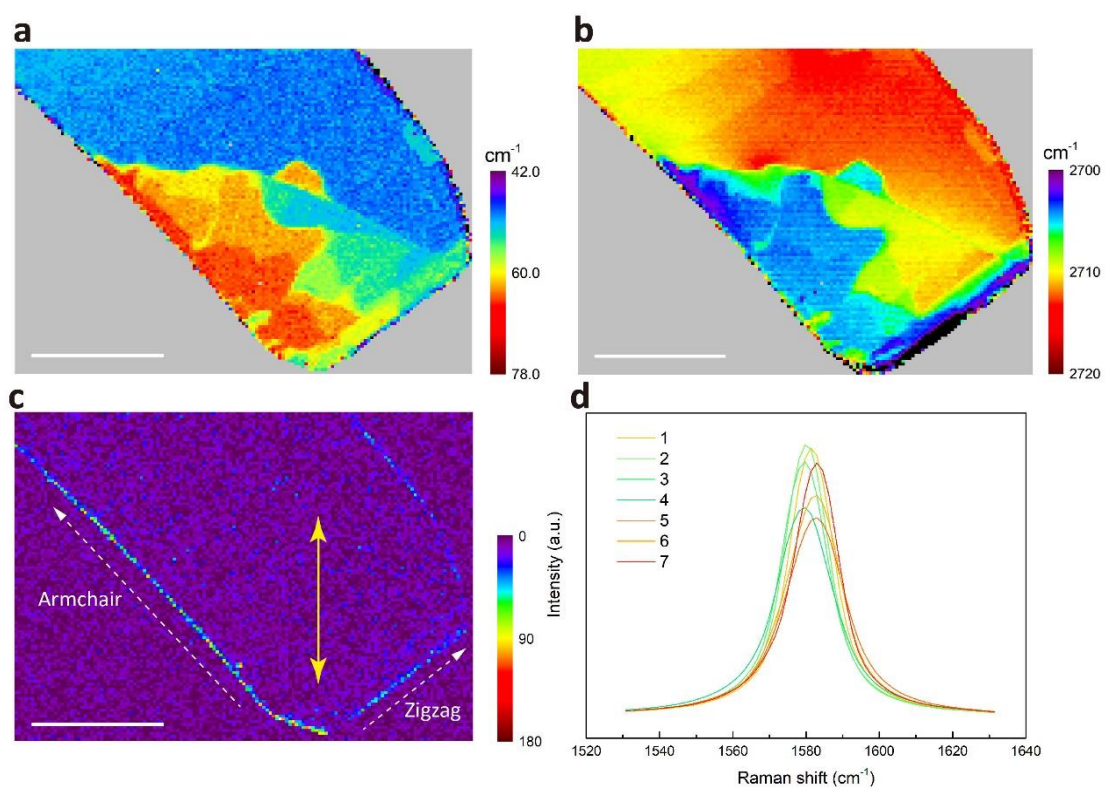

**Figure S2.** Raman maps of Flake2. (a) 2D band width map. (b) 2D band position map. (c) D band intensity map. The yellow arrow indicates the direction of laser polarization. The scale bars are all 20 μm. (d) G bands corresponding to the 1 to 7 domains shown in Figure 2d of the main text.

Apart from strain, the factors that also influence the line width and peak positions of G band and 2D band include the presence of doping or defects. The samples measured in our study are all freshly exfoliated flakes, which reduces the effect of doping. The D band intensity map (e.g., Figure S2c) indicates that our flakes are defect-free[10].

## 4. The stacking order of the domains in graphite films

The layer arrangements of the two allotropes of graphite films (≥ 3 layers) are shown in Figure S3a,b. We labelled the two sets of inequivalent atoms in each layer with A and B, representing the two sublattices of the hexagonal structure. For Bernal stacking, the atoms of the i+2 layer lie exactly on top of the $i^{th}$ layer, while for rhombohedral stacking, right above the centre of the hexagons in the $i^{th}$ layer, the atoms of the i+1 layer and i+2 layer are of inequivalent sites. In N-layer graphite film, as shown in Figure S3c, an alternating AB and BA stacking makes perfect Bernal stacking, while the repeated AB (or BA) stacking generates perfect rhombohedral stacking. For stacking faults, there are mixed segments of Bernal and rhombohedral stacking. In principle, for an N-layer graphite film there should be $2^{N-2}$ possible stacking sequences and the perfect ABA (or ABC) stacking corresponds to N-1 repeated Bernal (or rhombohedral) segments. The proportion and distribution of rhombohedral stacking segments in the stacking sequence of graphite film are the two important factors that affect its electronic structure, thus changing the Raman scattering response. For Flake2, the stacking orders of regions 1 and 7 should be perfect ABA and ABC stacking respectively. For regions 2 to 6, each of these domains has its specific proportion and distribution of Bernal and rhombohedral stacking segments.

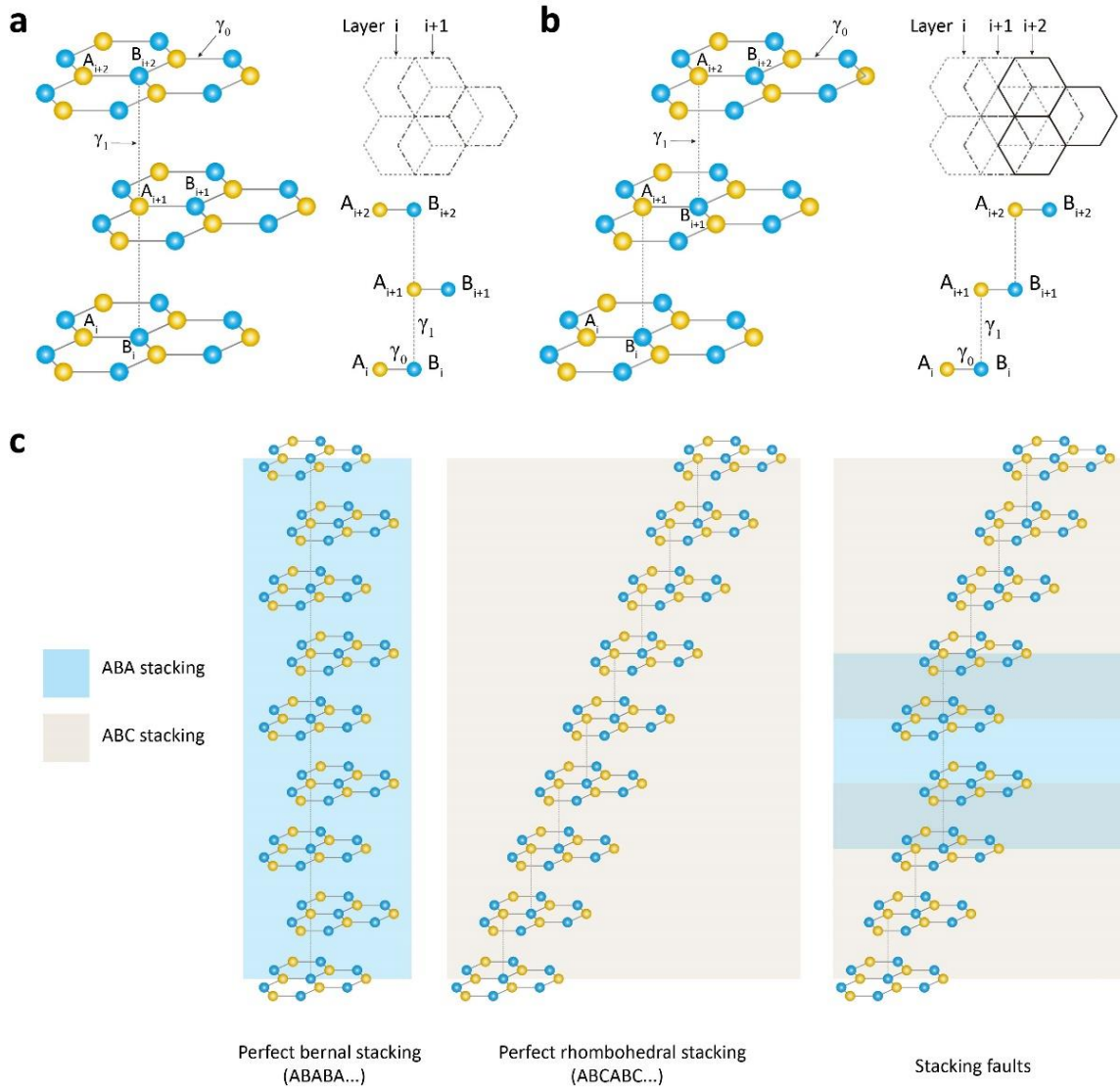

**Figure S3.** Atomic structures of ABA and ABC stacking and the schematic of stacking faults. (a,b) Atomic structures of Bernal stacking and rhombohedral stacking, i is the layer index. The blue and yellow spheres represent the A and B sublattices of the graphene honeycomb structure. In each panel, the top right figure is the top view and the bottom right figure is the side view of the unit cell. The hopping parameters $\gamma_0$ and $\gamma_1$ describe the nearest neighbour coupling within each layer and the coupling of the interlayer vertical bonds respectively. (c) Schematics of perfect ABA stacking, perfect ABC stacking and stacking faults with Bernal and rhombohedral stacking segments.

## 5. The distinct response of the domains in mechanical properties and near-field infrared plasmon reflection

We performed Quantitative Nanoscale Mechanical mode of AFM on another exfoliated graphite film (Flake6), as shown in Figure S4. The Raman map of the ratio of the integral area of 2*D* band ranging 2675-2705 cm$^{-1}$ to the area ranging 2705-2735 cm$^{-1}$ (Figure S4a) indicates the existence of different domains. In the AFM topography image (Figure S4b), the flake shows the homogeneous thickness and no domains are visible. Nevertheless, the mechanical channels of AFM (Figures S4c-f), which reflect the elastic modulus of the sample, exhibits several regions with distinctly different contrasts and matched shape and dimensions compared to the Raman map.

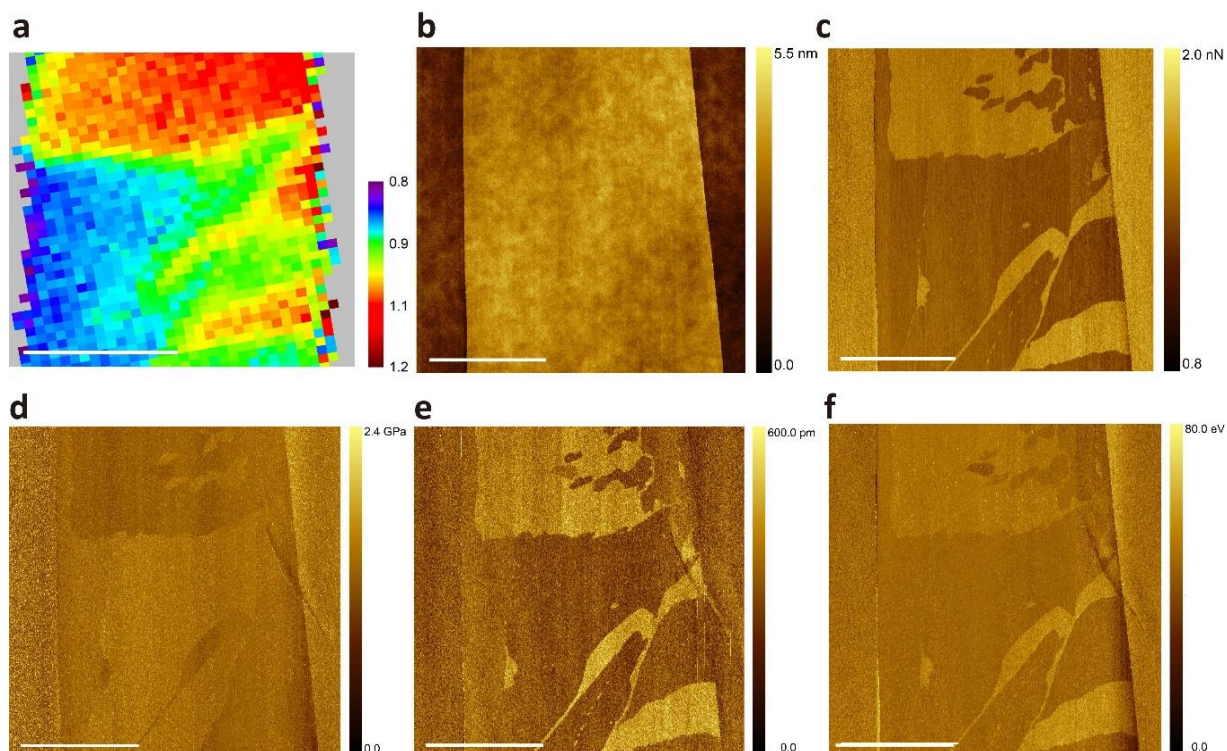

**Figure S4.** The correspondence between Raman map and mechanical response in AFM of graphite film with different domains. (a) Raman map of the ratio of the integral area of 2D band ranging 2675-2705 cm$^{-1}$ to that ranging 2705-2735 cm$^{-1}$. (b) AFM topography of Flake6. (c-f) The adhesion, Derjaguin-Muller-Toporov (DMT) modulus, deformation and dissipation channels of the flake taken at Quantitative Nanoscale Mechanical mode. The flake is ≈1.5 nm in thickness. The scale bars are 5 μm.

We performed near-field infrared nanoscopy on another exfoliated graphite film (Flake7) with different domains (Figure S5c). In the AFM topography image (Figure S5a), the flake shows the homogeneous thickness and no domains are visible whereas the near-field infrared nanoscopy image exhibits patterns which match those in Raman map.

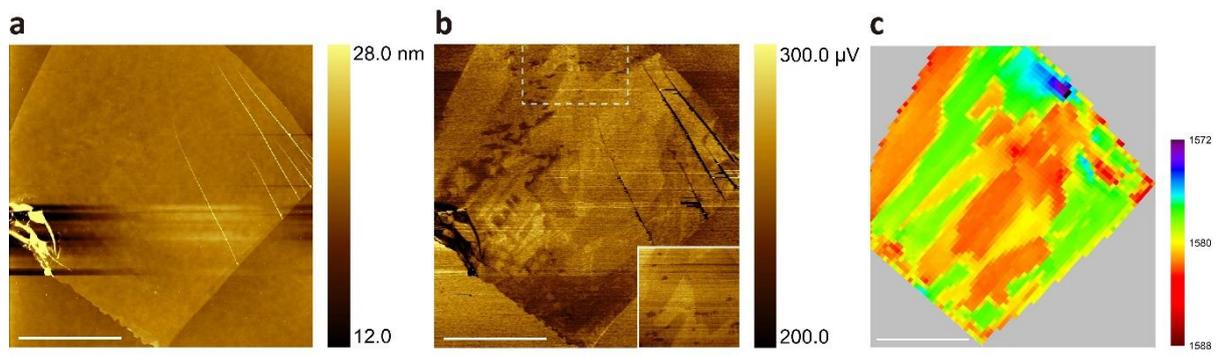

**Figure S5**. The correspondence between Raman map and near-field infrared nanoscopy image of graphite film with different domains. (a) AFM topography of Flake7. (b) The NF2-R channel of near-field infrared nanoscopy. (c) Raman map of the *G* band position. The flake is ≈1.5nm in thickness. The scale bars are 10 μm.

## 6. Edge chirality determination of graphite films by Raman spectroscopy

It is reported that D band in the Raman spectrum can be used to determine the chirality of the graphene edges[11-13]. A perfect zigzag edge cannot produce a D peak in the double resonance process, whereas the boundary conditions of the double resonance scattering process are only satisfied by the armchair edges[11, 14]. Thus the intensity of the D band near the armchair edge is always stronger than that near the zigzag edge[12, 14-16]. Another feature of D band is that its intensity shows a strong polarisation dependence: I(D) ∝ $\cos2\theta$ or $\cos4\theta$, where θ is the angle between laser polarisation and the flake edge[13, 16].

If the angle between two adjacent edges is 30° (modulo 60°), the two edges should be armchair and zigzag respectively. We selected the graphite flakes with these special angles (30°, 90°, or 150°) and then put them in a way that the direction of laser polarisation lies in the angular bisector of the angle between the adjacent edges, as shown in Figure S6, in order to diminish the influence of polarisation dependence of different edges. According to the different responses of edge chirality to the D band, the edge with a stronger D band intensity should be armchair, and the other should be zigzag. Even though the edges of graphite flakes are not always perfectly smooth, they are on average predominantly either zigzag or armchair in nature, allowing to determine the edge chirality by Raman characterisation.

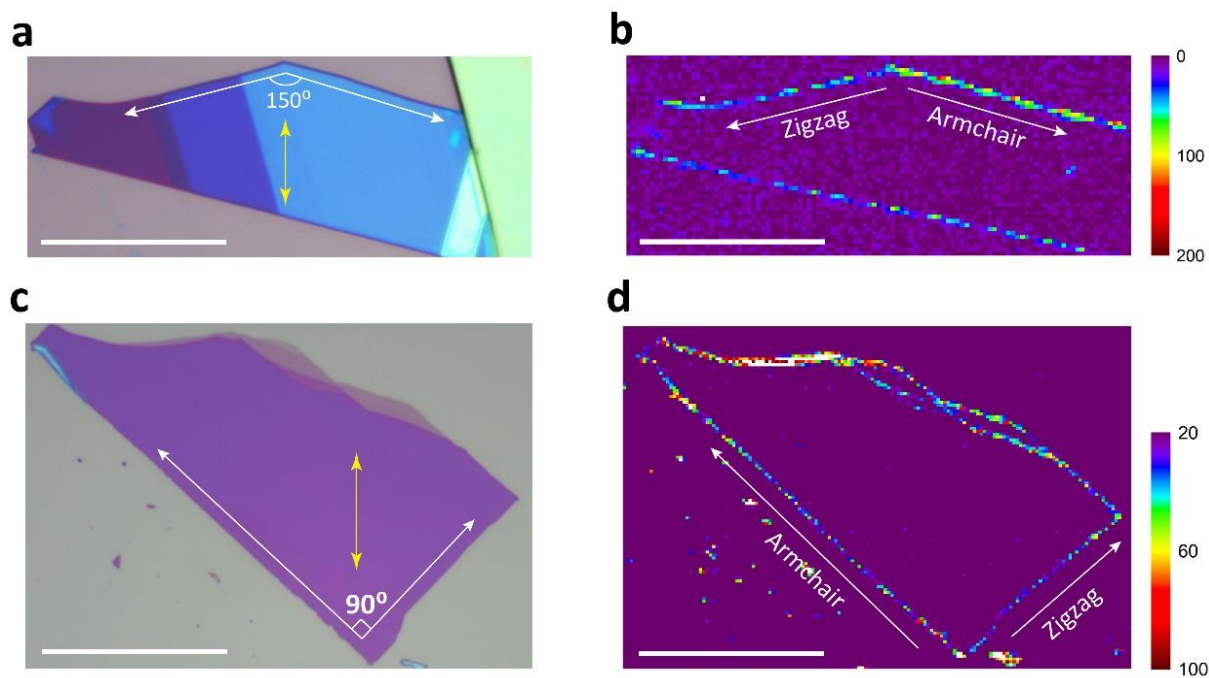

**Figure S6.** Edge chirality determination from Raman D band. (a) Optical micrograph of Flake3. (b) D band intensity map of Flake3. (c) Optical image of Flake4. (d) *D* band intensity map of Flake4. The yellow arrows in (a) and (c) indicate the direction of laser polarisation. The scale bars are 20 μm.

**Table S1.** Transformation of the stacking order of trilayer graphene along the displacement vector (d is the distance of energy favourable displacement, α is the C-C bond length, and n is an integer).

| Edge type along the dislocation vector | | | zigzag | armchair | |
|---|---|---|---|---|---|
| Origin stacking type | ABC | d | √3nα | (3n+1) α or (3n+2) α | 3nα |
| | | Final stacking type | ABC | ABA | ABC |
| | ABA | d | √3nα | 3nα | (3n+1)α or (3n+2)α |
| | | Final stacking type | ABA | ABA | ABC |

## 7. Consistency of the Raman signal measured by both sides of the graphite film

To confirm that the Raman signal comes from the entire graphite flake rather than the first few layers of the flake, we carried out the Raman spectroscopy on both sides of a suspended graphite flake (Flake8) with the thickness of around 9 nm (Figure S7b). The graphite flake was transferred onto a $MoS_2/Si_3N_4$ grid with the diameter of the holes being around 2 μm, Figure S7a. Figures S7c and S7d show the 2D band width maps of the suspended region measured while the top side and back side of the sample were facing upwards, respectively. The two Raman maps again show different stacking order in the suspended region among the flake and the distribution of the domains are consistent between the two maps, which means that the laser can penetrate through the flake with the thickness of at least 9nm.

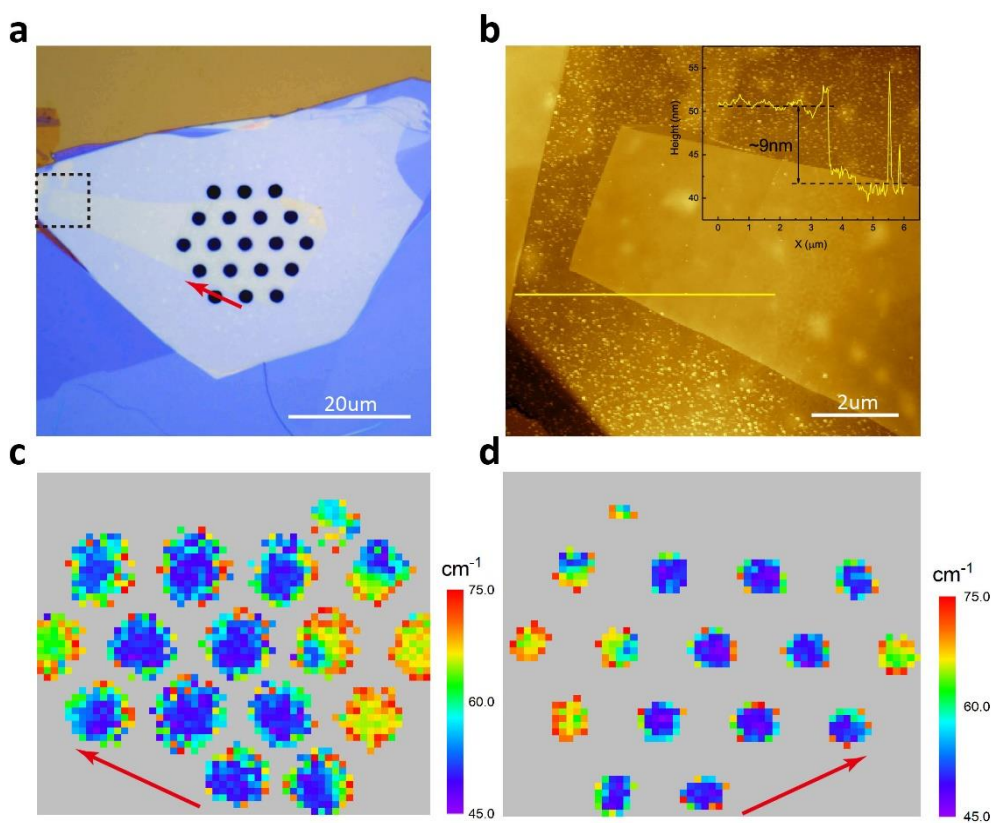

**Figure S7.** 2D band width Raman maps measured by both sides of the graphite film. (a) Optical micrograph of Flake8 suspended on $MoS_2/Si_3N_4$ grid. (b) AFM image of the dashed line region in (a) and height profile of the flake. (c) 2D width Raman maps of Flake5 with the top side facing upwards and (d) with the back side facing upwards. The red arrows in panels (a,c,d) indicate the layout of the holes covered by graphite film. Panels (c) and (d) are of mirror symmetry and the distribution of the domains in each panel match each other.

## 8. Calculation of the fraction of ABC stacking and the cross sectional TEM

To evaluate the fraction of ABC stacking, the intensity ratio (R) between the six inner-most first-order $\{10\bar{1}0\}$ spots and the six second-order $\{11\bar{2}0\}$ spots is firstly calculated by:

$$R = \frac{\sum_{i=1}^{6} I_i^{(1)}}{\sum_{i=1}^{6} I_i^{(2)}},$$

where $I_i^{(1)}$ and $I_i^{(2)}$ are the intensitiy of each of the first- and second-order peaks. The fraction of ABA stacking and ABC stacking can then be calculated by: [17]

$$F_{ABA} = \frac{R}{0.444} - 0.005,$$
$$F_{ABC} = 1 - F_{ABA}.$$

According to the method above, the calculated R for hole 1[#] and 2[#] are 0.02±0.01 and 0.46±0.01 respectively, giving the fraction of ABC for hole 1[#] and 2[#] as 96% and 2%, respectively.

For atomic-resolution cross-sectional S/TEM imaging, a dual-beam FIB/SEM instrument was used to prepare an electron transparent cross-sectional TEM specimen (see Methods section). The region of interest (ROI) contains both ABA and ABC stacked domains, confirmed by Raman mapping (black dashed region in Figure S8c). The ROI is then covered by BN flake as a protection layer, and located by correlating its optical image and Raman mapping (Figures S8a and c) with SEM imaging inside the FIB/SEM facility (Figure S8b) for cross-section specimen preparation. In the high-resolution STEM image (Figure S8d), individual atomic layers of the graphite flake are clearly resolved and the number of layers (9 layers for graphite) can be easily counted. As shown in Figure S8d, cross-sectional imaging reveals the existence of different stacking sequences in the selected region, which contains regions of both ABA and ABC stacking.

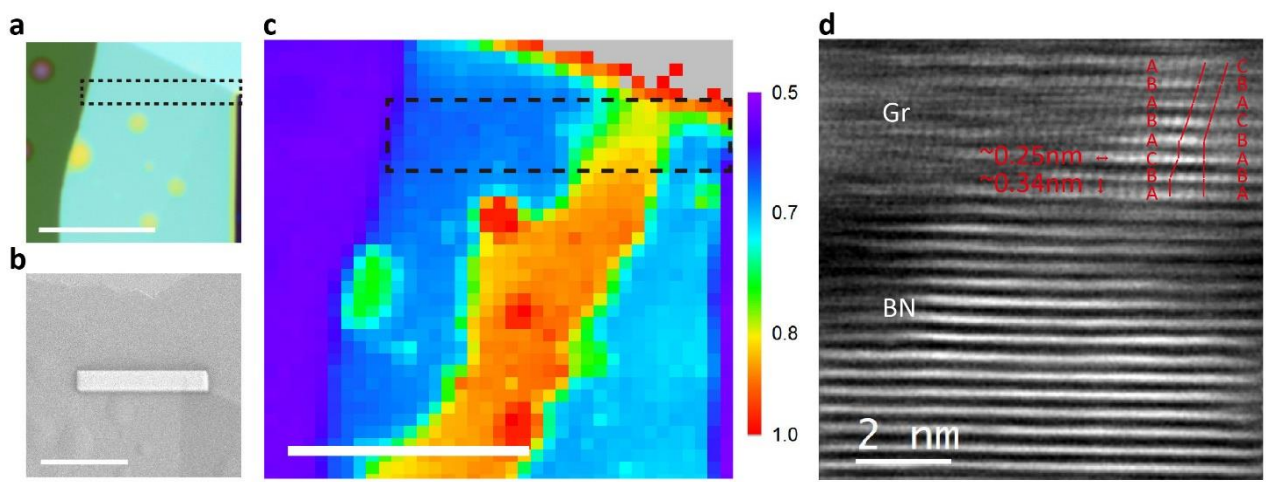

**Figure S8.** Cross-sectional STEM specimen preparation and imaging of a graphite flake containing both ABA and ABC domains. (a) Optical image taken from the flake on $SiO_2$/Si substrate, please note that the flake was covered by hBN to protect it during STEM specimen preparation. (b) Corresponding SEM image showing the region of interest in the FIB-SEM. (c) Corresponding 2D ratio Raman map of the flake. The scale bars of (a-c) are 10μm. (d) High-resolution STEM ADF image taken from the ABC/ABA graphite flake. The variation in the atomic positions indicates different basal plane stacking sequences are present in the graphite flake.

# References

1. Lui, C. H.; Malard, L. M.; Kim, S.; Lantz, G.; Laverge, F. E.; Saito, R.; Heinz, T. F. *Nano Lett* **2012,** 12, (11), 5539-44.
2. Blake, P.; Hill, E. W.; Castro Neto, A. H.; Novoselov, K. S.; Jiang, D.; Yang, R.; Booth, T. J.; Geim, A. K. *Appl. Phys. Lett.* **2007,** 91, (6), 063124.
3. Herziger, F.; May, P.; Maultzsch, J. *Physical Review B* **2012,** 85, (23).
4. Ivanov, D. Y.; Novoselov, K. S.; Dubrovskii, Y. V.; Sablikov, V. A.; Vdovin, E. E.; Khanin, Y. N.; Tulin, V. A.; Esteve, D.; Beaumont, S. *Phys. Low-Dimens. Struct.* **2000,** 3-4, 55-65.
5. Cong, C.; Yu, T.; Sato, K.; Shang, J.; Saito, R.; Dresselhaus, G. F.; Dresselhaus, M. S. *ACS Nano* **2011,** 5, (11), 8760-8.
6. Nguyen, T. A.; Lee, J. U.; Yoon, D.; Cheong, H. *Sci Rep* **2014,** 4, 4630.
7. Cong, C.; Yu, T.; Saito, R.; Dresselhaus, G. F.; Dresselhaus, M. S. *ACS Nano* **2011,** 5, (3), 1600-5.
8. Mafra, D. L.; Samsonidze, G.; Malard, L. M.; Elias, D. C.; Brant, J. C.; Plentz, F.; Alves, E. S.; Pimenta, M. A. *Phys. Rev. B* **2007,** 76, (23).
9. Maultzsch, J.; Reich, S.; Thomsen, C. *Phys. Rev. B* **2004,** 70, (15).
10. Eckmann, A.; Felten, A.; Mishchenko, A.; Britnell, L.; Krupke, R.; Novoselov, K. S.; Casiraghi, C. *Nano Lett.* **2012,** 12, (8), 3925-3930.
11. Ferrari, A. C.; Basko, D. M. *Nat Nanotechnol* **2013,** 8, (4), 235-46.
12. You, Y.; Ni, Z.; Yu, T.; Shen, Z. *Applied Physics Letters* **2008,** 93, (16), 163112.
13. Geim, A. K. *Science* **2009,** 324, (5934), 1530-1534.
14. Cancado, L. G.; Pimenta, M. A.; Neves, B. R.; Dantas, M. S.; Jorio, A. *Phys Rev Lett* **2004,** 93, (24), 247401.
15. Begliarbekov, M.; Sasaki, K.; Sul, O.; Yang, E. H.; Strauf, S. *Nano Lett* **2011,** 11, (11), 4874-8.
16. Zhang, Y. B.; Tan, Y. W.; Stormer, H. L.; Kim, P. *Nature* **2005,** 438, (7065), 201-204.
17. Latychevskaia, T.; Son, S.-K.; Yang, Y.; Chancellor, D.; Brown, M.; Ozdemir, S.; Madan, I.; Berruto, G.; Carbone, F.; Mishchenko, A.; Novoselov, K. S. *Frontiers of Physics* **2018,** 14, (1).